# Superlattice Patterns in Surface Waves

A. Kudrolli*, B. Pier, and J.P. Gollub+

*Department of Physics, Haverford College, Haverford PA 19041, USA, and*

*Physics Department, University of Pennsylvania, Philadelphia PA 19104, USA*

## Abstract

We report novel superlattice wave patterns at the interface of a fluid layer driven vertically. These patterns are described most naturally in terms of two interacting hexagonal sublattices. Two frequency forcing at very large aspect ratio is utilized in this work. A superlattice pattern ("superlattice-I") consisting of two hexagonal lattices oriented at a relative angle of 22º is obtained with a 6:7 ratio of forcing frequencies. Several theoretical approaches that may be useful in understanding this pattern have been proposed. In another example, the waves are fully described by two superimposed hexagonal lattices with a wavelength ratio of $\sqrt{3}$, oriented at a relative angle of 30º. The time dependence of this "superlattice-II" wave pattern is unusual. The instantaneous patterns reveal a time-periodic stripe modulation that breaks the 6-fold symmetry at any instant, but the stripes are absent in the time average. The instantaneous patterns are not simply amplitude modulations of the primary standing wave. A transition from the superlattice-II state to a 12-fold quasi-crystalline pattern is observed by changing the relative phase of the two forcing frequencies. Phase diagrams of the observed patterns (including superlattices, quasicrystalline patterns, ordinary hexagons, and squares) are obtained as a function of the amplitudes and relative phases of the driving accelerations.

# 1  Introduction

Surface waves produced by the Faraday instability are known to give rise to many different patterns (including stripes, squares, hexagons, and even quasicrystalline patterns) as a function of driving frequency and amplitude, viscosity, and the driving waveform [1-4].  However, regular space-filling patterns formed as a result of nonlinearity are even more diverse than the wide range reviewed by Cross and Hohenberg [5].  For example, in recent optical experiments in a Kerr-like medium, phase locking of several wave vectors results in novel patterns with several unequal wavevectors [6].  In this paper we report novel regular patterns observed in experiments on surface waves generated by two frequency forcing that extend earlier systematic work limited to single-frequency forcing [3].  *We will refer to these new patterns as superlattices because these are composed of two discrete but interacting sublattices.*  Their occurrence extends the striking variety of symmetric states than can occur in nonlinear surface waves.

In our earlier study hexagonal wave patterns were observed for low driving frequencies, for which the gravitational restoring force was comparable to the capillary force.  For high frequencies, i.e. in the capillary limit, square patterns were observed for low viscosity ($\nu < 50$ cm$^2$ s$^{-1}$), and textured stripes for higher viscosity.  Recent theoretical work by Zhang and Viñals has explained these observations using quasi-potential equations derived from the underlying hydrodynamic equations [7].

Good agreement with measured onset accelerations for two frequency forcing have been found with predictions of linear stability analysis of the full hydrodynamic equations [8].  Edwards and Fauve [4] observed 12-fold quasicrystalline patterns using two frequency forcing with ratio 4:5.  They argued that an even frequency perturbed by an odd frequency breaks the subharmonic symmetry (invariance with respect to translation in time by one driving period) and therefore allows quadratic terms (which would otherwise be excluded) to appear in the amplitude equations.  Stabilization of patterns such as hexagons can then occur by quadratic interaction as in non-Boussinesq convection.  These 12-fold quasicrystalline patterns may be described as two hexagonal lattices that are oriented at 30º.  A clear mechanism for their formation is not available, although recent work on a generalized Swift-Hohenberg model equation has shown 12-fold quasi-patterns [9].

In other work by Müller using 1:2 forcing, triangular and hexagonal lattices were observed; the type of lattice could be selected by using a third perturbing frequency [10], but no superlattices were reported. These experimental observations were reproduced theoretically by Zhang and Viñals using an extension of their quasi-potential equations to the case of two frequencies [11].

In the present work we report and discuss several new patterns formed with even-odd forcing. We study two frequency ratios (6:7 and 4:5), to explore the formation of patterns with novel symmetries. In both cases, there is a prominent region of hexagons in parameter space. By making relatively small adjustments in the relative amplitude or phase of the forcing components, we find several distinct patterns that are described most naturally as being composed of multiple hexagonal sublattices. In these patterns, the spatial power spectrum contains peaks at smaller wavenumbers than those observed at the onset of surface waves (for either driving frequency). These superlattice patterns arise from an instability of the base hexagonal lattice that is formed at onset. Superlattice structures are of course common in condensed matter physics, but to our knowledge the term has not previously been applied to patterns in nonlinear systems.

## 2 Experimental setup and forcing function

The apparatus is essentially as described in Ref. [8]. The experimental setup consists of a 32 cm diameter circular aluminum container filled with silicone oil to a height of 3 mm. This material gives stable behavior over many weeks, and is available over a wide range of kinematic viscosity, though the present work is limited mainly to $v = 20\text{-}50$ cm$^2$ s$^{-1}$ because of the large number of other parameters that need to be varied. The fluid depth is generally greater than the viscous penetration depth at the typical fluid oscillation frequency. The container is rigidly attached to a Vibration Test Systems electromagnetic shaker that is capable of applying peak forces of 2200 N.

The forcing waveform that controls the acceleration of the container is described by

$$a(t) = a(\cos(\chi)\cos(n\omega t) + \sin(\chi)\cos(m\omega t + \phi)) , \qquad (1)$$

where, $\omega = 2\pi f$, $n$ and $m$ are integers, $1/f$ is the overall period $T$ of the driving, and $\chi$ is used to control the relative amplitudes. By measuring the actual acceleration and using feedback, the acceleration is forced to follow Eq. (1) to within about 1%. We generally choose $n$ to be even and $m$ to be odd so that nonlinear interactions can occur at quadratic order, as explained in Sec. 1. We thoroughly explore two cases: $(m,n)=(6,7)$ and $(4,5)$, and we systematically vary $a$, $\phi$, $\chi$ for each case. For some purposes it is useful to define $a_n = a\cos(\chi)$ and $a_m = a\sin(\chi)$.

Lighting is provided by a circular array of lights, and the reflected light from the fluid surface is imaged. Roughly speaking, nodes appear dark because light is deflected away from the camera, while antinodes appear bright. The intensity is a (nonlinear) function of the surface slope, reaching a maximum at an angle (4.2º) that is often small compared to the slopes typically present in the patterns. We typically average over 1-2 wave periods; this causes the apparent wavelength to be half the actual wavelength, since there are two nodes (or antinodes) per period. Though the images are not quantitative measures of the surface height function or slope, they are useful for determining the symmetries of the patterns. In some cases, instantaneous images are used to determine the variation of the waves with time within one wave period.

The patterns formed at the surface are imaged using a 512 x 512 pixel Dalsa variable scan CCD camera controlled by the same computer that generates the driving waveform of the shaker. The camera exposure times may be varied with a minimum exposure time of about $10^{-3}$ s. The phase of the image acquisition with respect to the forcing function is adjustable by means of a programmable liquid crystal shutter. Pattern analysis is implemented using Khoros image processing software.

## 3 Experimental results

### 3.1. Parameter space for $6\omega$:$7\omega$ forcing

First we discuss the patterns obtained using $n=6$ and $m=7$, i.e. a 6:7 frequency ratio. We find that in the capillary limit (where the capillary force dominates over gravitation) the patterns do not depend strongly on the frequency $f$. Therefore detailed studies of the patterns observed as a function of $a$, $\chi$, and $\phi$ in Eq. (1) were conducted at $f = 16.44$ Hz. The phase diagram of the patterns obtained as a function of the relative strength of acceleration is shown in Fig 1(a) for $f = 16.44$ Hz, and $\phi = 20º$. The data shown in the phase diagram were obtained by observing the pattern after incrementing the acceleration $a$ in steps of 0.1g, and $\chi$ and $\phi$ in steps of 5º.

A square lattice is observed when the accelerations $a_6$ and $a_7$ are substantially unequal. This observation is consistent with previous studies using single frequency forcing at the same viscosity [3]. The square patterns corresponding to the even frequency are harmonic since the phase of the fluid motion has the same sign after a drive period $T$. On the other hand, the square patterns corresponding to the odd frequency are subharmonic, i.e. they change sign after one drive period and will have the same sign only after two periods. (However the subharmonic pattern does recur at a shorter time interval $2T/7$. The use of the term subharmonic refers to the phase of the pattern after one drive period $T$.)

More complex patterns are obtained when both components are comparable. In this situation, hexagons are formed at onset, as shown in Fig. 1. This transition is subcritical, since hysteresis is observed at onset. Because of the observation of hysteresis and the role played by the odd frequency $7\omega$, these hexagons are thought to arise from quadratic interactions in the amplitude equations [4]. They arise from a fundamentally different mechanism than those observed with single frequency forcing, where the interactions are at cubic order in the amplitude equations and triad resonances (which occur at low frequencies) are important [3, 12]. As the variable $\chi$ (which specifies the relative strength of the two frequencies) is increased near onset, the wavelength changes suddenly from the one corresponding to the even forcing term to that corresponding to the odd term at a particular value $\chi_o \equiv 61.5º$ that is independent of $\phi$. This point in the parameter space may be regarded as a "bi-critical point".

Time-dependent patterns are also visible in the phase diagrams. These include: Transverse Amplitude Modulations (TAM), which have been predicted in a simpler form [13] and observed [14] previously; and spatiotemporal chaos (STC) similar to that which develops from square patterns for single frequency forcing [3].

The main features of the phase diagram are relatively independent of $\phi$. This is shown in the phase diagram near the bi-critical point for fixed $\chi$ as shown in Fig 1(b). However, the secondary instability leading to disorder does depend significantly on $\phi$.

### 3.2. Superlattice-I patterns and their spectra

An additional instability occurs as the acceleration is increased in the vicinity of the bi-critical point (see Fig 1(a)). We refer to the resulting pattern, shown in Fig. 2, as a *superlattice-I (SL-I) pattern* (to distinguish it from another case that we describe later in this section). This pattern shows a triangular (3-fold) lattice at large scales, with each lattice point being composed of discrete clusters of three small cells. Each of the small cells is approximately the same size as the hexagons in the "nearby" hexagonal state. We use stroboscopic illumination to determine that this pattern is harmonic with the drive period $T$; the image shown in Fig. 2 has been obtained with an exposure time equal to the drive period.

We compare the two-dimensional power spectrum of the SL-I state with that of the hexagonal pattern in Fig. 3. The SL-I spectrum has the six-fold symmetry of a hexagonal lattice, but the minimum lattice vector (corresponding to the *inner* set of six peaks) is quite different in magnitude from the wavenumber $k_o$ that is forced directly and corresponds to the onset hexagons, shown in Fig. 3(b). Instead, $k_o$ corresponds to the *fourth* circle in Fig 3(a) that is concentric with the origin; these peaks have been highlighted artificially and are *farther* from the origin by a factor $\sqrt{7}$. Furthermore, the 12 peaks on this circle are *not* equally spaced. They may be described as two sets of 6 peaks forming two hexagons oriented at an angle $2 \cdot \sin^{-1}(1/2\sqrt{7}) \approx 22°$ to each other. (This situation is quite different from the 12 peaks of the quasicrystalline pattern discussed in Sec. 3.3. In that case, there are also 12 dominant peaks, but they are equally spaced on a circle centered at the origin.)

We note that the onset wavenumber $k_o$ corresponds (via the dispersion relation) to the $6\omega$ forcing. We propose that the SL-I pattern results from resonances involving quadratic interactions between the highlighted peaks on the fourth circle with wavenumber $k_o$; such interactions are allowed for harmonic patterns produced by even-odd forcing, and they are capable of generating all of the other peaks in the power spectrum. The 6 peaks located on the slightly larger *fifth* circle have a wavenumber that approximately corresponds to the $7\omega$ term. This frequency matching may also be important in generating the SL-I pattern [9], but we are unable to assess its importance.

*Phenomenological description of the SL-1 state:* As we have indicated, a description of this SL-1 state can be given in terms of two interacting hexagonal lattices; we give it more explicitly here. Each hexagonal lattice may be specified by an instantaneous surface height function given (to within a constant of proportionality) by

$$F_{hex}(x,y) = \sum_{i=1}^{3} \cos(\mathbf{k_i} \cdot \mathbf{r} + \beta_i) \, , \text{ with} \tag{2}$$

$\mathbf{k_1} = (1,0)$; $\mathbf{k_2} = (-1/2, \sqrt{3}/2)$; $\mathbf{k_3} = (-1/2, -\sqrt{3}/2)$; $\mathbf{r} = (x,y)$.

We denote the arguments of the cosine functions by $\psi_i$; the total phase $\Phi \equiv \Sigma \psi_i = \Sigma \beta_i$ can take on only two possible values: 0 (corresponding to the centers of the hexagons having positive or upward displacement) and $\pi$ (corresponding to hexagons with centers down at the given instant). Hexagonal patterns have points of 6-fold symmetry where $\psi_i = 0$ (for all *i*), and points of triangular symmetry for which $\psi_i = +2\pi/3$ (for all *i*) or $\psi_i = -2\pi/3$ (for all *i*).

If a second lattice, rotated by an angle $\theta = 2 \cdot \sin^{-1}(1/2\sqrt{7}) \approx 22°$, is also present, then the combined pattern is *periodic* because of the commensurability or resonance condition $2\mathbf{k_1'} - \mathbf{k_3'} = \mathbf{k_1} - 2\mathbf{k_3}$. However, it is still necessary to specify the total phases and the relative positions of the centers of the hexagons of the two lattices. We choose the total phases of the two lattices to be the same, i.e. $\Phi = \Phi' = 0$. (If they are unequal, the constructed pattern does not resemble the experimental pattern.) The texture also depends on the displacement of the second lattice with with respect to the first one. It turns out that the pattern we observe is obtained if a point of 6-fold symmetry of the first lattice (where all the $\psi_i = 0$) coincides with a point of 3-fold symmetry of the second lattice (where all the $\psi_i = \pm 2\pi/3$). This pattern is shown in Fig. 4(a). Another interesting pattern is obtained by superimposing *either* points of hexagonal symmetry $\psi_i = 0$, or points of triangular symmetry of different type, i.e. $\psi_i = 2\pi/3$ with $\psi_i = \pm 2\pi/3$. This pattern is shown in Fig. 4(b); symmetry considerations have been used to argue for its stability [15], but we do not observe it. Finally, if points on the two lattices with no rotational symmetry are superimposed, then stripe patterns are generally obtained.

From these considerations, we learn that to form the observed SL-1 pattern from two hexagonal lattices, it is essential *not only* that the wavevectors of the two lattices be locked at the correct angle in Fourier space, but *also* that a phase-locking condition be satisfied in real space: the positions of the two patterns must have the relationship described in the preceding paragraph. Note that the constructed pattern of Fig. 4(a) closely resembles the experimental one even though it does not contain the smaller wavevectors that are present in the power spectrum of the experimental image. These smaller wavevectors are probably a mixture of (a) nonlinear interactions arising from three-wave resonances between the various Fourier components and (b) imaging nonlinearity. There is no easy way to determine the relative importance of these two contributions.

*Transition to the SL-1 state:* The inner ring of peaks in Fig. 3(a) is present in the SL-I state but not in the hexagon state. Therefore we use the strength of these peaks to follow the transition qualitatively as the acceleration is varied, while remaining cognizant of the fact that imaging nonlinearity can contribute significantly to their strength. We define a *superlattice amplitude* $S_I$ by first integrating the power spectrum azimuthally and then integrating over $\Delta k = 0.3$ cm$^{-1}$ centered at the peak corresponding to the first ring of peaks. The variation of $S_I$ with driving acceleration *a* is shown in Fig. 5. The contribution from the background noise has been subtracted. The transition appears to be continuous, and could be a transcritical bifurcation. Visually the domains of the SL-I state spread with increasing acceleration until they gradually cover the entire container. However, it is possible that the transition is actually discontinuous, and that the discontinuity is masked by a slightly inhomogeneous driving acceleration (variation ± 1.5%).

*Defects:* The superlattice-I patterns generally show weak time dependence due to the presence of defects in the patterns. The most prominent of these are phase defects that causes the "triangular" structure in the SL-I pattern to vary locally in orientation. A typical defect-free region is of the same size as that shown in Fig. 2. Small domains of ordinary hexagons are also present with accompanying grain boundaries. We found that the number of defects does not decrease appreciably with time.

*Other superpositions:* Edwards and Fauve [4] have reported quasicrystalline patterns with 6:7 forcing frequency ratio. We did not observe them with 6:7 forcing, but our viscosity was significantly lower (20 cm$^2$ s$^{-1}$ versus 100 cm$^2$ s$^{-1}$ in their case). Given the large number of parameters that affect the superlattice patterns, it was impractical to explore the variation with viscosity in the present work. Earlier work in our laboratory using single frequency forcing showed that the pattern symmetry depends on the viscosity [3].

### 3.3. Parameter space for $4\omega:5\omega$ forcing

Next we discuss the patterns observed with a 4:5 forcing frequency ratio. The phase diagram of the patterns obtained as a function of the two forcing amplitudes is shown in Fig. 6(a) for $f = 22$ Hz, and $\phi = 16^o$. As for the 6:7 frequency ratio, the frequency again is high enough that the waves are in the capillary limit. The patterns do not depend strongly on $f$. The overall structure of the phase diagram is similar to that observed for the 6:7 frequency ratio: square lattices when one component is much larger than the other, and hexagons and superlattices when both components are comparable. The bi-critical point is also at a similar location in the phase space: $\chi_o \equiv 61.5^o$ and is independent of $\phi$. As in the 6:7 frequency ratio case, the main features of the phase diagram do not change qualitatively with $\phi$ except near $\chi_o$.

Quasicrystalline patterns are observed as the acceleration is increased beyond onset, for $\chi$ slightly less than $\chi_o$. An example of such a 12-fold quasicrystalline pattern is shown in Fig 7. The field of view is approximately 20 cm x 20 cm. Competition between hexagons and quasicrystalline patterns, which makes them time-dependent, is observed for parameters between those of the pure states. An example of this competition is shown in Fig. 8. We believe that this competition is inherent and not due to an inhomogeneous driving force because the domains exchange position in time.

We studied the transition from hexagons to quasicrystalline patterns using a spectral *quasicrystalline amplitude* $S_Q(a)$ that is similar that used for the SL-I state; it denotes the amplitude of the inner ring of peaks in the corresponding spectrum; they arise due to nonlinear interactions between the main spectral components (but are also affected by imaging nonlinearity) and is shown in Fig. 9. The onset of hexagonal-quasicrystal competition appears to be abrupt; this regime leads smoothly to the pure quasicrystalline state around $a/g$=8.9.

### 3.4. Superlattice-II state

When $\phi$ is increased with $\chi \approx 61^O$ (slightly less than $\chi_o$), the quasicrystalline patterns are replaced by a superlattice structure that is quite different from the superlattice-I state discussed in Sec. 3.2. This superlattice-II pattern, whose stable region is shown in Fig.6(b), is subharmonic with respect to $f$ and shows a distinctive periodic time dependence. It was first reported by Pier in his undergraduate thesis [16]. An image obtained by averaging over two drive cycles is shown in Fig. 10 at two different scales. This pattern is composed of hexagonal cells, but there is also a larger wavelength hexagonal lattice superimposed upon it. The cells in the large lattice have higher amplitude and hence appear darker (since more light is scattered away from the camera). The higher amplitude of these cells was checked with a strobe light using side illumination; it is clearly real, and not a lighting artifact.

Additional information about this pattern may be obtained by examining the two-dimensional power spectrum, which is shown in Fig. 11. The peaks corresponding to the hexagonal patterns formed at onset are indicated by the outer ring of six circles. The corresponding wavenumber $k_O$ is that obtained by using the dispersion relation with $2\omega$ as the wave frequency (the subharmonic of the $4\omega$ forcing). In addition to these peaks and their harmonics, a set of 6 peaks occurs at smaller $k$. The wavenumber of this inner ring is smaller than $k_O$ by a factor $\sqrt{3}$.

The experimental spectrum suggests a description of the superlattice-II state in terms of two hexagonal lattices oriented at 30º to each other, and with wavenumbers having a ratio equal to $\sqrt{3}$. (Using the inner peaks alone could reproduce all of the observed spectral peaks by quadratic interactions, but only at the cost of not including the directly forced modes in the description.) A simulated pattern obtained in this way is shown in Fig. 12. Each of the two component lattices has been chosen to place points of hexagonal symmetry at the origin so that $\psi_i = 0$ (for all $i$) in the notation defined below Eq. (2). The result is qualitatively similar to the experimental pattern of Fig. 10(a). In this description, we have ignored any direct forcing at the wavenumber corresponding to $5\omega$, though it may be significant.

The SL-II state displays additional complexity that is not shown by the SL-I state. An instantaneous image (exposure over 1/20th of the drive period) reveals this complexity: the 6-fold symmetry is broken, and the observed pattern depends on the instant at which the image is obtained. Examples of instantaneous images obtained at four different phases (with respect to $4\omega$) separated by $T/20$ are shown in Fig. 13. A stripe modulation is visible that is not present in the average images. Power spectra of these images indicate that the wavenumber of this modulation is half that of the onset hexagons. The stripes are not a simple amplitude modulation of the primary standing wave, since in that case they would survive in the averaged picture as do the subharmonic stripes observed with single frequency forcing [3]. The stripe modulation is always present; time averaging yields the hexagonal superlattice-II patterns shown in Fig. 10.

The phase diagram as a function of $\phi$ is shown in Fig. 6(b). The onset acceleration for this pattern does not change appreciably with $\phi$. A disordered region occurs for parameters between the quasicrystalline SL-II states.

## 4 Discussion

We have reported novel superlattice patterns that occur when Faraday waves are driven at two frequencies. They are closely related to the hexagonal state that occurs near onset. The superlattice-I state, shown in Fig. 2, is harmonic with respect to the total drive frequency $\omega$. It can be described in terms of two superimposed hexagonal lattices whose wavenumbers are both equal to that of the onset hexagonal state, but whose wave*vectors* are oriented at an angle of $2\sin^{-1}(1/2\sqrt{7}) \approx 22°$ (Fig. 3). (While other representations are possible, we believe it is important to include at least one of the wavevectors that are directly forced.) The discussion of Sec. 3.2 shows that the two lattices are phase-locked together in real space so that a point of 6-fold symmetry of the first lattice coincides with a point of triangular symmetry of the second lattice. The new SL-I state differs from the other regular patterns, and from the quasicrystalline patterns which also have 12 primary spectral peaks, in that the overall orientational symmetry (3-fold) is less than the number of primary spectral components.

This SL-I pattern seems to be one of the generic possibilities that can be expected on the basis of symmetry considerations, as explained by Silber and Proctor [17]. These authors show that a low-dimensional model with degenerate bifurcations can be constructed to reproduce the observed transition sequence. The transition shown by their model is hysteretic, whereas the experimental transition appears to be continuous. However, in studying the SL-I state and the transition connecting it to the hexagonal state, we are limited by imaging nonlinearity and a slightly inhomogeneous driving acceleration. An alternate approach to understanding patterns resulting from forcing at two frequencies, such as the SL-I pattern, has been proposed by Lifshitz and Petrich [9]. It is based on a Swift-Hohenberg model with two preferred length scales and both quadratic and cubic nonlinear terms in the wave amplitude. The quadratic term includes the effect of triad interactions between standing waves. In this model, the selected patterns do more than satisfy symmetry considerations; they lead to a lower value of a certain Lyapunov functional. We are unable to test this model, but it is not inconsistent with what we observe.

Using a $4\omega{:}5\omega$ frequency ratio, we find and characterize a second superlattice state, which we call superlattice-II (Fig. 10). Its time average can be represented primarily as a combination of two hexagonal lattices differing in wavenumber by a factor $\sqrt{3}$. This state shows a remarkable time-dependent stripe modulation (Fig. 13) that breaks the hexagonal symmetry at an instant, yet leaves this symmetry unbroken on average. There is at present no theory applicable to this state.

The various hexagonal, quasi-crystalline, and superlattice states show additional complexity due to the presence of defects and competition between patterns that are adjacent in parameter space. Competition between hexagonal and quasicrystalline patterns was illustrated in Fig. 8. Using spectral methods, we find that the onset of this competition is abrupt. In the regime of hexagons near onset, hepta-penta defects maybe formed by sudden increase of driving acceleration. These generally anneal out over a very long time-scale and a defect free pattern is usually obtained.

A "clean" hexagonal pattern is best obtained by slowly increasing the acceleration near the wave onset. In this respect the hexagonal patterns obtained using two frequency forcing differ from those obtained for single frequency forcing, where defects, including π-phase defects are more common (see Ref. [3]). No phase defects were observed for two frequency forcing over the range of parameters we investigated. The ease with which defects are eliminated may be related to the fact that the transition from the flat state is subcritical, i.e., there is a small amount of hysteresis, typically about 0.1g in the driving amplitude $a$. This hysteresis was first discussed by Edwards and Fauve [4]. This situation is different from that of single frequency forcing, where the coupling giving rise to hexagons is at third order in the amplitudes, and both gravity and capillarity are significant.

Defects are also observed in the superlattice-I state. These defects cause the *orientations* of the triangular structures shown in Fig. 2 to vary from place to place. (This variation can be seen when larger areas are examined.) In fact we were unable to obtain a single superlattice-I pattern in which the whole pattern had the same orientation. These defects move slowly, and therefore the patterns are weakly time-dependent. On the other hand the superlattice-II patterns show few defects when the acceleration is increased slowly. Defects that do appear in the superlattice-II state can probably be attributed to the slightly inhomogeneous driving acceleration.

It seems likely that other patterns with distinct symmetries can also be created using two-frequency forcing; because of the large number of parameters that can be varied, we have not explored all of the possibilities. We believe that the patterns discussed in this paper can be used to explore the role of nonlinearity in stabilizing structures with different symmetries that are composed of interacting waves.

## Acknowledgments

We appreciate helpful discussions with M. Abraham, R. Lifshitz, and Mary Silber. W.S. Edwards designed the apparatus, taught us about two-frequency forcing, and was involved in the early stages of this investigation. B. Boyes provided technical support. This work was supported in part by the National Science Foundation under Grants DMR-9319973 and DMR-9704301.

# Figure Captions

Fig. 1: Phase diagrams of patterns obtained for a forcing frequency ratio of 6:7 (see Eq. (1)). (a), the axes represent the forcing amplitudes $a_6$ and $a_7$ at the two frequencies $6f$ and $7f$, and the phase angle $\phi$ is held fixed at $16^o$ ($\nu = 20$ cm$^2$ s$^{-1}$, $f$=16.4 Hz). (b) Phase variation of the patterns for fixed $\chi = 61^o$. H=hexagons; $S_s$=subharmonic squares; $S_h$=harmonic squares; SL-I=Superlattice-I state; TAM=transverse amplitude modulations; STC=spatiotemporal chaos. In shaded regions competition occurs.

Fig. 2: Example of the superlattice-I pattern obtained for two-frequency forcing with ratio 6:7 ($f$ = 16.44 Hz, $\phi = 20^o$, $\chi = 61^o$).

Fig. 3: Two-dimensional power spectrum of the superlattice-I state (a) and hexagonal state (b). The two sets of circles in black highlight the peaks corresponding to the two hexagonal lattices of the same wavenumber as the onset hexagons shown in (b).

Fig. 4: Patterns formed by superposing two hexagonal lattices at an angle $\theta = 2 \cdot \sin^{-1}(1/2\sqrt{7}) \approx 22^o$. The total phases defined below Eq. (2) are the same for both lattices. (a) A point of 6-fold symmetry of the first lattice coincides with a point of triangular symmetry of the second lattice; this gives a pattern that closely matches the experimental SL-I pattern with overall triangular symmetry. (b) A point of 6-fold symmetry of one lattice corresponds with a similar point of the other lattice.

Fig. 5: The superlattice amplitude $S_I(a)$ (see text) as a function of acceleration shows a continuous transition from hexagons to the SL-I state ($f$ = 16.44 Hz, $\phi = 20^o$, $\chi = 61^o$).

Fig. 6: Phase diagrams of patterns obtained for a forcing frequency ratio of 4:5 with $f$=22 Hz. The axes are the same as in Fig. 1. SL-II=superlattice-II state (see text); $S_s$=subharmonic squares; $S_h$=harmonic squares; H=hexagons; Q=quasicrystalline patterns; x=competition between quasipatterns and hexagons; TAM=transverse amplitude modulations; STC=spatiotemporal chaos; dark shading=no stable pattern near onset; light shading=competition between neighboring states.

Fig. 7: Example of a quasicrystalline pattern obtained at $\phi = 16^o$ ($f$ = 22 Hz, 4:5 frequency ratio). This pattern has 12-fold orientational symmetry but no translational order.

Fig. 8: Quasicrystalline patterns and hexagons compete on a slow timescale in the region between the pure states indicated by the symbol x in the phase diagram of Fig 6(a). (a,b) are separated in time by 1600 drive periods.

Fig. 9: The quasicrystal spectral amplitude $S_Q$ (*a*) as a function of acceleration shows a discontinuous jump at the onset of the region of quasicrystal-hexagon competition ($f = 22$ Hz, $\phi = 16^o$, $\chi = 61^o$).

Fig. 10: (a) Superlattice-II state obtained with 4:5 forcing frequency ratio by changing $\phi$ to $60^o$. The wave periodicity is $2T$; this pattern was obtained by averaging the image over two drive periods ($f = 22$ Hz). (b) Pattern after magnification by a factor of 4. The scale is within a few percent of that used in Fig. 2.

Fig. 11: Two dimensional power spectrum corresponding to Fig. 10. The outer ring of 6 peaks (marked by circles) are those of the hexagonal pattern formed near onset. The inner ring of highlighted peaks are associated with the transformation of the hexagonal state into the superlattice state. Their wavenumber is a factor $\sqrt{3}$ smaller than that of the onset hexagons.

Fig. 12: Simulated SL-II superlattice pattern generated by adding two hexagonal lattices with a wavenumber ratio $\sqrt{3}$; the orientations of the lattices differ by $30^o$.

Fig. 13: The six fold symmetry of the SL-II averaged pattern is broken by the presence of a temporal modulation. Instantaneous images obtained with exposure times of 1/20th of the drive period $T$ show a stripe modulation during the drive cycle. Starting times are as follows: (a) t = 6 $T$ / 20, (b) t = 7 $T$ / 20, (c) t = 8 $T$ / 20, and (d) t = 9 $T$ / 20.